# A comparative study of paired versus unpaired deep learning methods for physically enhancing digital rock image resolution

Yufu Niu[1], Samuel J. Jackson[2], Naif Alqahtani[1], Peyman Mostaghimi[1] and Ryan T. Armstrong[1]

[1]School of Minerals and Energy Resources Engineering, The University of New South Wales, NSW 2052, Australia.

[2]CSIRO Energy, Clayton South, VIC, Australia

## Abstract

X-ray micro-computed tomography (micro-CT) has been widely leveraged to characterise pore-scale geometry in subsurface porous rock. Recent developments in super resolution (SR) methods using deep learning allow the digital enhancement of low resolution (LR) images over large spatial scales, creating SR images comparable to the high resolution (HR) ground truth. This circumvents traditional resolution and field-of-view trade-offs. An outstanding issue is the use of paired (registered) LR and HR data, which is often required in the training step of such methods but is difficult to obtain. In this work, we rigorously compare two different state-of-the-art SR deep learning techniques, using both paired and unpaired data, with like-for-like ground truth data. The first approach requires paired images to train a convolutional neural network (CNN) while the second approach uses unpaired images to train a generative adversarial network (GAN). The two approaches are compared using a micro-CT carbonate rock sample with complicated micro-porous textures. We implemented various image based and numerical verifications and experimental validation to quantitatively evaluate the physical accuracy and sensitivities of the two methods. Our quantitative results show that unpaired GAN approach can reconstruct super-resolution images as precise as paired CNN method, with comparable training times and dataset requirement. This unlocks new applications for micro-CT image enhancement using unpaired deep learning methods; image registration is no longer needed during the data processing stage. Decoupled images from data storage platforms can be exploited more efficiently to train networks for SR digital rock applications. This opens up a new pathway for various applications of multi-scale flow simulation in heterogeneous porous media.

# 1.Introduction

With the aid of X-ray micro-computed tomography (micro-CT), the pore-scale structure of subsurface rock can be accurately characterised to understand how fluids flow in porous rock for enhanced oil recovery [1], carbon dioxide sequestration [2], multiphase flow in fuel cells [3], and various other applications. Based on the standard workflow [4], digital rock grey-scale images are initially obtained from micro-CT and then segmented into two phases – pore and grain for direct numerical simulation or pore network modelling [5]. The achievements of those applications, however, are highly dependent on how accurate the pore-scale geometries are captured from the micro-CT images. Ideally, there is a trade-off between image resolution and field-of-view (FOV), high-resolution images more accurately depict pore geometries at the cost of reducing the FOV. In contrast, low-resolution images have larger FOV but cannot represent the true structural details of the rock. This presents a critical challenge since high-resolution data can depict pore characterisation precisely while FOV needs to be large enough to represent the presence of heterogeneity [6] [7].

In the past few decades, super-resolution (SR) technology has been applied to circumvent the trade-off between resolution and FOV. SR aims to reconstruct a high resolution (HR) counterpart of a degraded low resolution (LR) image [8]. Traditional SR methods have been demonstrated to improve image resolution, such as stochastic approaches [9] [10], Bayesian method [11], neighbour embedding [12] [13], sparse representation [14] [15], projection onto convex sets (POCS) approach [16], and example-based approach [17]. These traditional methods, however, have their own drawbacks. For instance, neighbour embedding does not implement well on complicated images with textural regions [13]. POCSs and example-based methods need high computational time [17] [18]. Sparse representation has the challenge of balancing the relations between dictionary size and computational cost [15].

Recent advances in deep learning have exceeded traditional methods to solve the single image super resolution task (SISR) using convolutional neural networks (CNN) or generative adversarial networks (GAN) based on image quality metrics, such as Peak Signal to Noise Ratio (PSNR) and Structural Similarity Index Measure (SSIM). Dong et al. [19] firstly developed a deep convolutional network, called SRCNN, by learning the end-to-end mapping between bicubic LR and HR data. Thereafter, more advanced deep neural networks have been proposed for SISR inspired by SRCNN using various effective structures. Dong et al. [20] first introduced a fast super-resolution convolutional neural network (FSRCNN) using normal deconvolution layers, which can reduce the computational time. However, the deconvolutional layer can cause redundancies during the upsampling procedure [21]. Instead of using the deconvolution layer, an efficient sub-pixel convolutional neural network (ESPCNN) was proposed to learn the upscaling process for SISR by rearranging the feature maps of the low-resolution image to high-resolution image mapping [22]. Thereafter, more neural network oriented approaches were presented for SISR, such as VDSR [23], DRCN [24], EDSR [25], SRDenseNet [26], MemNet [27], WDSR [28], and so forth. Most of the current deep learning models need paired training data, which is not always available. Therefore, researchers have applied various generative adversarial network (GAN) approaches to solve SR problems using unpaired training data, such as SRGAN [29], CinCGAN [30], High-to-Low

GAN [31], DSR/CSR [32], and others. Paired algorithms usually provide higher accuracy based on PSNR/SSIM while unpaired algorithms are more flexible to leverage for real world data [31] [30].

In digital rock physics, SR techniques can provide large-scale domains at high resolution for flow simulation where the large FOV can represent heterogeneous features [33]. Recent SR studies on digital rock images demonstrated that CNN-based SR models can generate high-quality images [34] [35]. These previous works evaluated SR performance based on grey scale analyses, e.g., histogram data, differential maps, as well as image quality metrics. These standards, however, cannot explicitly determine the physical accuracy of the SR images for petrophysical analyses, such as porosity, absolute/relative permeability, which are critical parameters for digital rock physics. Wang et al. [34] demonstrated the permeability of SR images can be consistent with their HR ground truth (GT) counterparts with various segmentation thresholds using paired SRCNN and SRGAN methods. Niu et al. [36] further illustrated that the physical accuracy of SR encompassing porosity, permeability, pore size distribution, and Euler characteristic was equivalent to its HR counterpart using an unpaired CinCGAN. A validation work by [33] demonstrated the reliability and efficiency of EDSR on the application of multiphase flow simulation on large scale heterogeneous porous media. Results show that the physical accuracy of EDSR results is comparable to the related GT data and experimental data.

The paired and unpaired SR deep learning methods raise an important question. Can unpaired methods achieve equivalent physical accuracy when compared to paired method? In this paper, we examined two state-of-the-art SR paired/unpaired deep learning models – EDSR (paired) and CinCGAN (unpaired) to enhance the image resolution of an imaged carbonate sample, which includes resolved and sub-resolved pores that are challenging to characterise from a single resolution image. In general, both EDSR and CinCGAN are found to precisely capture the edge sharpness and high frequency texture of the SR grey-scale images, which cannot be resolved in LR images. Simulated petrophysical properties using pore network modelling show that both EDSR and CinCGAN can accurately reconstruct SR images comparable to their HR counterpart. Our results suggest that unpaired deep learning models can become an alternative way to enhance digital rock image resolution when paired data are unavailable. Image registration can be skipped to accelerate the entire image processing workflow. In addition, images from data archives can be exploited in unprecedented ways by using unpaired approaches to provide SR solutions.

## 2. Materials and Methods

### 2.1 Materials

A 6 mm heterogeneous Middle Eastern Carbonate (MEC) core cylindrical plug was initially scanned at LR (10.72 µm) and HR (2.68 µm) with scale factor of 4x. The imaging details are presented in Table 1. Basic settings, such as voltage, tube current, and exposure time are the same for both scans while the distance from the source determines the image resolution. The original 16 bits micro-CT images for this study can be found on digital rock portal. (https://www.digitalrocksportal.org/projects/362)

*Table 1: The scan details on LR HR MEC sample implemented by HeliScan Micro-CT facility at University of New South Wales* [37]

| Sample Name | Voxel Size (μm) | Voltage (kV) | Tube current (μA) | Distance from source (mm) | Exposure time (sec) | Scan duration (Hrs) | No. of Projection |
|---|---|---|---|---|---|---|---|
| HR MEC | 2.68 | 80 | 85 | 5.8 | 0.64 | 10.5 h | 2520 |
| LR MEC | 10.72 | 80 | 85 | 23.2 | 0.64 | 5.1 h | 2520 |

The original 3D 16 bits Micro-CT LR/HR images were initially registered using Avizo software. The images were then cropped to 380x380x1025 voxels for LR and 1520x1520x4100 voxels for the corresponding HR to remove the background. Afterwards, the 16 bits images were converted to 8 bits using standard image normalization as

$$\bar{p} = \frac{p - p_{min}}{p_{max} - p_{min}}, \qquad Eq.1$$

where $p$ is the grey scale value, while $p_{min}$ and $p_{max}$ are the maximum and minimum values by eliminating the extremums.

## 2.2 EDSR

EDSR was introduced by Lim et al. [25] as a 2D multi-scale CNN-based deep learning framework for SISR image enhancement. They utilised interpolated images, as inputs while GT images are the coupled HR images. EDSR encompasses two convolutional layers, a series of residual blocks and upsampling blocks for resolution enhancement [38]. In this paper, we extend the EDSR model to 3D, as shown in Fig.1(a). To alleviate the computational burden, we reduce the filter numbers in the convolutional layers from 64 to 32. Instead of inputting interpolated LR images, we utilised natural LR/HR images as inputs/outputs to retain the original image information. We also apply a trilinear upsampling method for resolution enhancement where the feature maps have the same scale as the output to replace the pixel shuffle upsampling method in the original EDSR. To train the EDSR model, the L1 loss function was applied to optimise the weights and biases.

$$L1 = \sum_{i=1}^{n} |y_{gt} - y_{predicted}|, \qquad Eq.2$$

where $y_{gt}$ is the ground truth data and $y_{predicted}$ is the predicted data from the neural network.

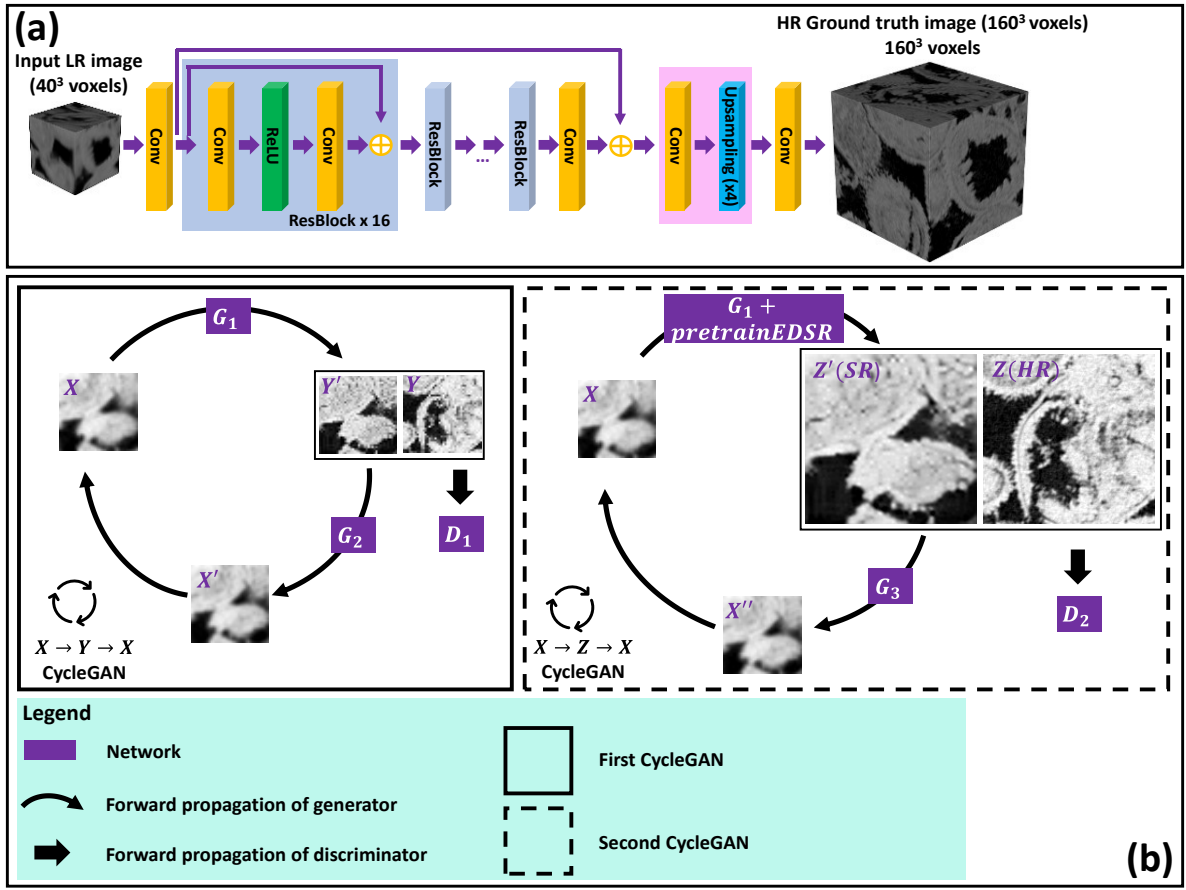

Fig.1: An overview of the architectures of the proposed deep learning models: (a) 3D EDSR, and (b) CinCGAN.

## 2.3 CinCGAN

GAN proposed by Goodfellow et al. [39] has been broadly applied in computer vision tasks, e.g., image segmentation [40] [41], SR [36], image denoising [42] [43], and image synthesis and manipulation [44] [45]. Among the applications of GANs, one is called Cycle-consistent GAN (CycleGAN) that was initially designed for image-to-image translation [46]. Yuan et al. [30] presented an unpaired CinCGAN SR model, which can generate high-quality SR images when compared with paired SR methods.

Fig.1 (b) shows the architecture of CinCGAN where two CycleGANs are applied to construct CinCGAN. To train a CinCGAN, the three datasets shown in Fig.1 (b) are required: (1) a low-resolution image ($X$), (2) a bicubic low-resolution image ($Y$) that is interpolated from ($Z$), and (3) a high-resolution image ($Z$). The first CycleGAN mapping is denoted as $X \to Y \to X$ in the black box of Fig.1 (b). $G_1$ generates fake image $Y'$ that is similar to the clean bicubic LR image ($Y$) in order to confuse the Discriminator, $D_1$. $G_1$ acts as a deblurring filter to clean the LR image ($X$) by regarding the bicubic LR ($Y$) as a reference. This is because the bicubic LR image ($Y$) is noise-free when compared with the input LR ($X$). Generator $G_2$ maintains a reverse mapping $Y \to X$ to reinforce the under-constrained mapping of $X \to Y$. Discriminator $D_1$ aims to distinguish the fake image $Y'$ generated by $G_1$ from the real image $Y$. The second

CycleGAN mapping is denoted as $X \to Z \to X$ in the black dotted box of Fig.1 (b). In this step, a pretrained 2D $EDSR$ model is initially trained between the bicubic LR image ($Y$) and HR image ($Z$). Then the trained $G_1$ from first CycleGAN and pretrained $EDSR$ are regarded as a new Generator ($G_1 + EDSR$) to generate a fake SR image $Z'$ similar to the real HR image ($Z$). Similar to the first CycleGAN, Generator $G_3$ adds an inverse downscaling mapping $Z \to X$ to constrain the solution. Discriminator $D_2$ aims to differentiate the fake SR image $Z'$ from the real HR image ($Z$). The loss function to optimize the weights and biases is

$$L_{Total}^{LR} = L_{GAN}^{total} + \lambda_1 L_{cycle}^{X-Y} + \lambda_2 L_{identity}^{X-Y} + \lambda_3 L_{TV}^{X-Y}$$
$$+ \beta_1 L_{cycle}^{X-Z} + \beta_2 L_{identity}^{X-Z} + \beta_3 L_{TV}^{X-Z}, \qquad Eq.3$$

where $L_{GAN}^{total}$ is total generator-adversarial loss, $L_{cycle}^{X-Y}/L_{cycle}^{X-Z}$ is cycle consistency loss, $L_{identity}^{X-Y}/L_{identity}^{X-Z}$ is identity loss, $L_{TV}^{X-Y}/L_{TV}^{X-Z}$ is total variation loss, $\lambda_1$, $\lambda_2$ and $\lambda_3$ are the weights for the different losses in the first $X \to Y \to X$ CycleGAN and $\beta_1$, $\beta_2$ and $\beta_3$ are the weights for the losses in second $X \to Z \to X$ CycleGAN. The details for each loss function can be found in the Supplemental Material [47].

**2.4 Mercury Intrusion Capillary Pressure**

To quantify the porosity of the macro- and micropores system of the tested carbonate sample, a Mercury Intrusion Capillary Pressure (MICP) test was conducted. The test was run using POREMASTER® by Quantachrome instruments on another sample from the same block [37]. The results were analysed using a suite of Thomeer hyperbolas [48]. Thomeer hyperbolas can be used to decode different pore systems through type-curve matching and superposition in porous media [49] [50]. A Thomeer hyperbola can be expressed as

$$\frac{B_v}{B^\infty} = \exp[-G/Log(P_c/P_d)], \qquad Eq.4$$

where $B_v$ is the volume of mercury injected, $B^\infty$ is the percentage of bulk volume intruded with mercury at infinite pressure, $G$ is a pore geometrical factor, $P_c$ is injection pressure (capillary pressure), and $P_d$ is the displacement pressure required for mercury intrusion to the largest pore throat. The Thomeer hyperbola parameters are depicted in Fig.2 (a). The related Thomeer hyperbolas matched to the experimental MICP data in Fig.2 (b) show a total porosity of 29.79% where macroporosity and microporosity account for 17.84% and 10.97% of the sample bulk volume, respectively.

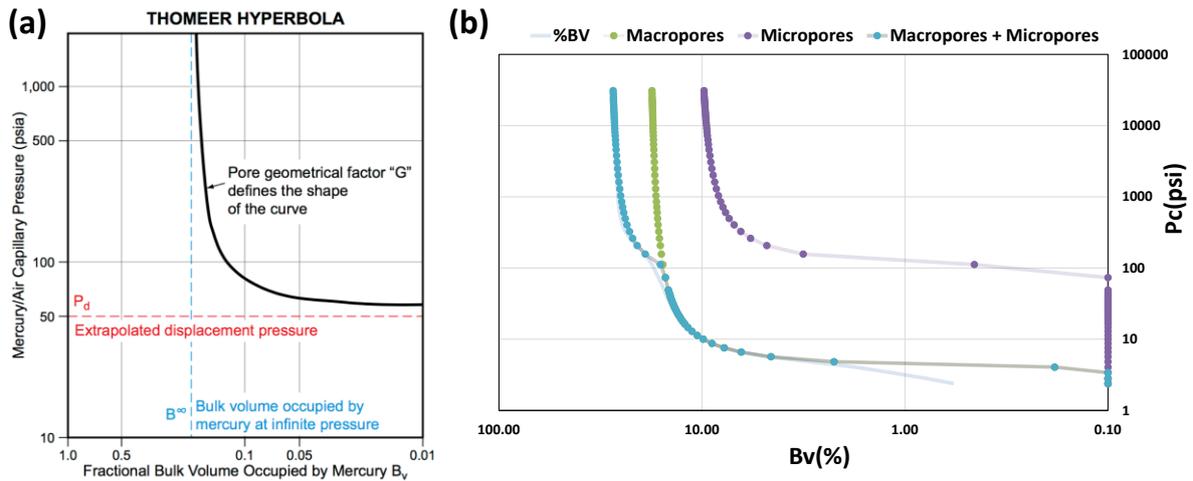

Fig.2(a): Thomeer hyperbola parameters used for characterizing each pore system in the carbonate sample [49] (b) Thomeer hyperbolas matched with the experimental MICP data [37].

### 2.5. Training Process

EDSR training data were extracted using a sequence of image patches with $40^3$ voxels for LR and $160^3$ voxels for the corresponding HR using a sliding window moving on the training/testing region with overlapping step sizes of 20 and 80 for LR and HR data, respectively. All LR data are coupled to the HR data. In total, there were 2080 and 512 image patches for training and testing, respectively. The example image patches for LR/HR ground truth images are shown in Fig.1(a). The Adam optimisation method was applied to update weights/bias in the EDSR [51]. The learning rate was initially set at $10^{-4}$ and decreased tenfold every twenty epochs. Batch size was 6 to reduce computational cost and 100 epochs were used for training.

The same number of image patches were generated for CinCGAN - $40^2$ voxels for LR, $160^2$ voxels for HR. The LR/ HR image patches were independently extracted from different FOVs with overlapping step sizes of 40 and 160 for LR and HR data. We initially trained the first $X \rightarrow Y \rightarrow X$ CycleGAN for 100 epochs to restore the noisy input data to clean data. Then, a pretrained EDSR was trained as an upscaling model for the second $X \rightarrow Z \rightarrow X$ CycleGAN training. The pretrained EDSR was trained using HR images and corresponding bicubic downsampled images. With the help of the pretrained EDSR, we load the trained $G_1$ from the first CycleGAN along with the pretrained EDSR and train the second $X \rightarrow Z \rightarrow X$ CycleGAN for another 50 epochs. All training was implemented with Adam optimisation [51]. Batch size was set to 8 and the initial learning rate was $10^{-4}$, then halved every twenty epochs.

All training was conducted using a NVIDIA GeForce RTX 2080Ti GPU. All code was developed using the PyTorch platform.

## 2.6. Validation

Fig. 3 depicts the workflow for validation of the reconstructed super-resolution images and quantitative petrophysical analyses. Steps 1 through 4 are explained in this section while Steps 5 through 9 are covered in the Results and Discussion section.

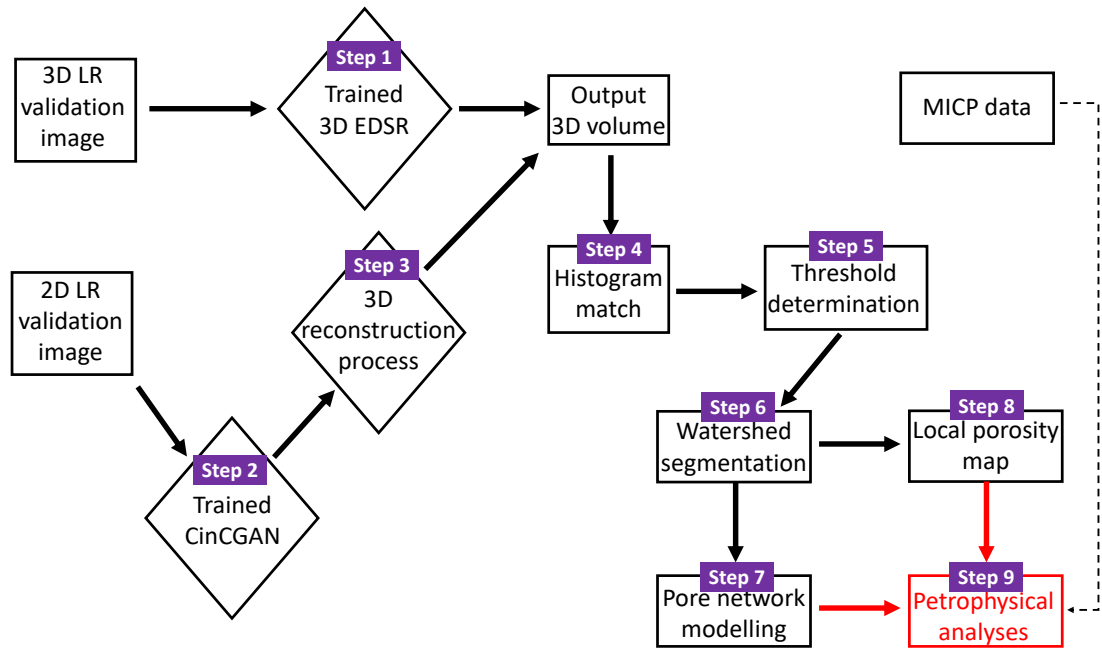

Fig.3: The overall workflow for validation of the reconstructed super-resolution images and petrophysical analyses.

When the training procedure was completed, a validation 3D LR volume (380x380x512 voxels) that has never been seen by the EDSR nor CinCGAN was fed into the pre-trained models to provide a corresponding 4x SR volume, i.e., 1520x1520x2048 voxels. The 3D EDSR model cannot input such a large 3D volume directly due to the GPU memory limitation. Herein, we split the validation volume into a series of sub-volumes (380x380x4 voxels) in the Z-axis direction. Each LR sub-volume was reconstructed and then stacked to form a full 3D SR validation image (1520x1520x2048 voxels). We visually observed inconsistent artefacts at the boundaries between sub volumes in z plane shown in S1(a)-(b) in supplementary material [52] [52]. This is caused by padding of convolutional kernels since there is no prior information at image boundaries. However, S1(c) in supplementary material [53] shows that the inconsistent artefacts at image boundaries will not result in segmentation errors.

Running a 3D CinCGAN directly is time-consuming and overloads the internal memory of the GPU. Therefore, we implemented a few simple steps as demonstrated by our previous work to reconstruct 3D image using the 2D CinCGAN [36].

- We first input 512 2D LR images ($380^2$ voxels) in the X-Y plane to CinCGAN and reconstructed 512 2D SR images ($1520^2$ voxels) in the X-Y plane.
- Then, the 512 2D SR images ($1520^2$ voxels) in X-Y dimension are downsampled to 1520 2D LR images (380x512 voxels) in the X-Z plane using bicubic interpolation.

- The 1520 2D LR images (380x512 voxels) in the X-Z plane are then fed into the pretrained EDSR model to generate the final 3D SR volume (1520x1520x2048 voxels).

The procedure does not cause any coupling problems caused by using the two different networks [36]. This is because both the CinCGAN and pretrained EDSR model are trained using the same training data. S2(a)-(b) in supplemental materials [54] shows that there are no visually apparent inconsistent artefacts at image boundaries in z plane by the bicubic interpolation methods. The corresponding segmentation from S2(c) in supplemental materials [55] also shows that there are no boundary artefacts can affect the accuracy of segmentation.

To ensure that the SR results can be fairly compared with the HR ground truth, a histogram match method was implemented on the SR validation images using 'imhistmatch' function in MATLAB. The 'imhistmatch' function aims to adjust the histogram of the SR image to the HR ground truth reference image.

## 2.7 Pore Network Modelling

We use the conventional PNM approach presented (and available online) in [56] [57], which are updated versions of the original algorithms [58] [59]. Full details of the approaches can be found from references therein. Further validations of the PNM are available in [60] [61] [33] .

In summary, we use a maximal spheres algorithm to assign pore-bodies and throats to represent the pore space. The pore bodies and throats are then assigned shape-factors based on their geometry, and quasi-static capillary dominated drainage flow is simulated across the network, for a constant capillary pressure. At each capillary pressure equilibrium stage, single or multiphase transport (hydraulic or electric) can be simulated. Local conductivities are found either analytically or through empirical relationships, e.g., for corner-flow. Pore-body potential is solved for the network by enforcing conservation of flux at each pore body. These potentials can be averaged at the inlet and outlet, which when combined with corresponding fluxes can be used to obtain macroscopic transport properties, e.g., permeability, relative permeability, and formation factor. A water/decane system was utilised as fluid properties on our PNM simulation. The microporosity is not considered in the PNM.

## 3.Results and Discussion

In this section, we first visually observe the reconstructed grey-scale images, measure the resulting PSNR/SSIM, and report the computational performance of the SR algorithms. We then present an objective means for data segmentation for macroporosity and microporosity determination. Microporosity maps are generated and quantified in terms of void fraction and heterogeneity. Lastly, typical petrophysical properties are evaluated using pore network modelling. Overall, we provide a robust quantitative assessment of the resulting SR images in comparison to HR ground truth images and MICP data.

## 3.1 Reconstructed Images

Fig.4 shows the four validation volumes – LR, registered HR ground truth (HR-GT), EDSR validation with histogram match (EDSR-HM), and CinCGAN validation with histogram match (CinCGAN-HM). The images demonstrate the finer features that are captured in the HR and SR images. In addition, all images are within a similar grey-scale range, which will be important for image segmentation and subsequent evaluation of physical accuracy, which is qualitative evident in Fig.4 but also observed in the image histograms that will be presented in Section 3.2.

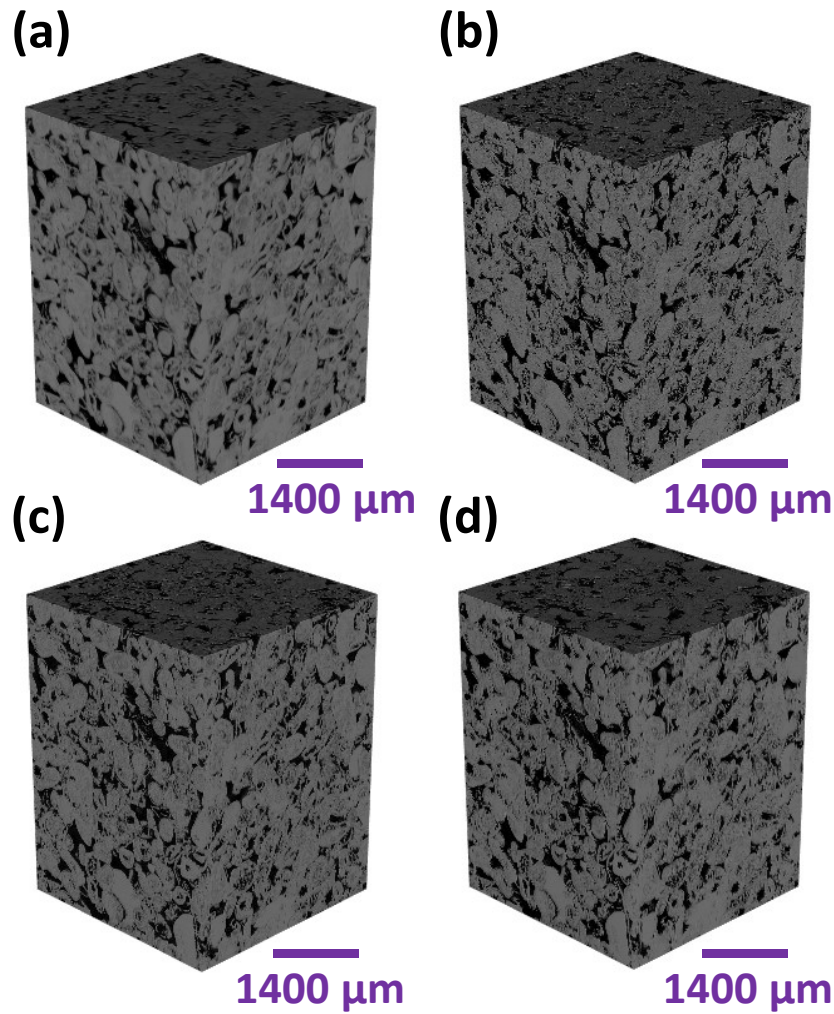

Fig.4: 3D rendering of the related validation volume for petrophysical analyse. (a) LR (380x380x512 voxels), (b) registered HR-GT (1520x1520x2048 voxels), (c) EDSR-HM (1520x1520x2048 voxels), (d) CinCGAN-HM (1520x1520x2048 voxels).

Table 2 provides a performance comparison of the EDSR and CinCGAN networks. CinCGAN needs less time to train but more time for SR reconstruction than EDSR. Overall, the total computational time of 2D CinCGAN remains lower than EDSR. In addition, both PSNR and SSIM of EDSR are 15.93% and 35.04% higher than CinCGAN respectively. This is because EDSR as a CNN-based method can immediately learn the mapping between LR and HR data using paired data while unpaired CinCGAN as a GAN-based approach causes more uncertainty when generating fake data from the learned distribution.

*Table 2: EDSR vs CinCGAN performance comparison - computational cost, PSNR and SSIM vs HR.*

|  | EDSR | CinCGAN |
|---|---|---|
| Training Data | Paired | Unparied |
| Total Training Time (mins) | 358 | 267 |
| Total Reconstruction Time (mins) | 8.5 | 17.2 |
| PSNR vs. HR | 16.81 | 14.50 |
| SSIM vs. HR | 0.370 | 0.274 |

Fig.5(a)-(d) shows 2D grey scale images for the LR, HR-GT, EDSR-HM, and CinCGAN-HM images. From global visual inspection, EDSR-HM and CinCGAN-HM can capture most of the texture details of the HR-GT image. When we look into the local regions shown in Fig. 5(e)-(h), we can see discrepancies between the images. The LR image displayed in Fig.5(e) does not capture the grey-scale textures of the microporous regions well. Also, the edges between the grain and macropores lack sharpness. The blue boxes in Fig.5(f)-(h) shows detailed differences between the HR-GT, EDSR-HM, and CinCGAN-HM images. EDSR-HM image shown in Fig.5(g) restores the high-frequency information when comparing with the HR-GT image in Fig.5(f). In contrast, CinCGAN-HM displayed in Fig.5(h) creates what appears to be unrealistic grey-scale textures. Fig.5(i) displays the grey-scale intensities along the line segments for the HR-GT, EDSR-HM, and CinCGAN-HM images. The locations of the line segments are depicted as the dashed lines in Fig.5(b)-(d). From the line profiles, we observe that the variation of the grey-scale values for the EDSR-HM image is similar to what is observed from the CinCGAN-HM image. The gradients for the line profiles at the interfaces between the macropores and grain are also similar for all images.

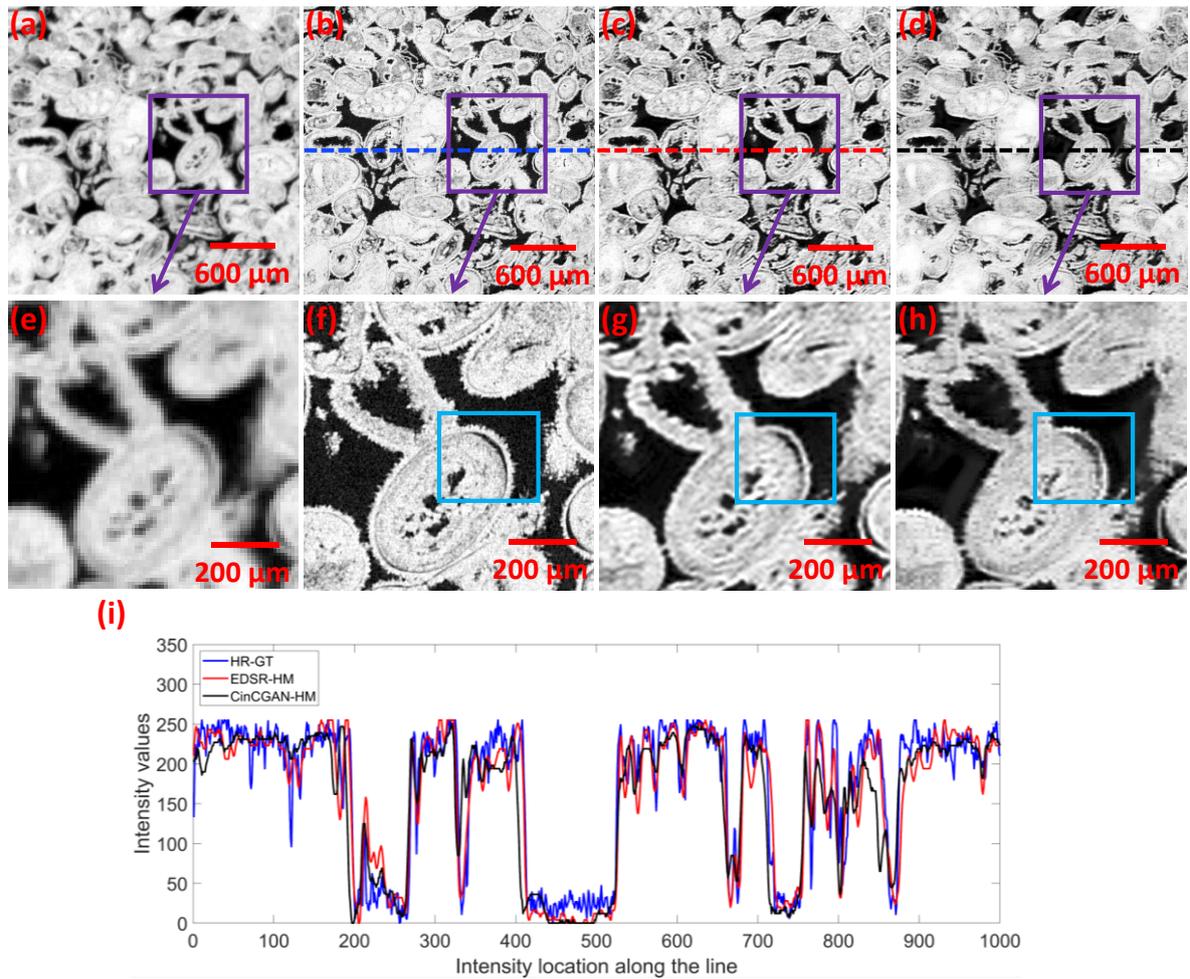

Fig.5: 2D Grey scale images and related image line profiles. (a) LR, (b) HR-GT, (c) EDSR-HM, (d) CinCGAN-HM, (e) amplified LR, (f) amplified HR-GT, (g) amplified EDSR-HM, (h) amplified CinCGAN-HM, (i) intensity cross-sections along line segments located on grey scale images for HR-GT, EDSR-HM and CinCGAN-HM.

**3.2 Image Segmentation**

The watershed-based method was applied to segment the image volumes [62]. To implement watershed segmentation, we initially defined two markers for macro-pore and solid phases. Then morphological watershed transformation algorithm [63] was applied for interphase region growing. In general, the micropores in carbonate rock are also called sub-resolution pores [64] which cannot resolved from the micro-CT images due to resolution limitations. The micropores can be defined by image resolution. For example, the image resolution of our high-resolution carbonate data is 2.68 µm which means that any pore diameters less than 2.68 µm can be regarded as micropores. Two regions (macropores and grains) were segmented initially. The microporosity is then defined within the 'grain' phase as a subsequent step. The main challenge for segmentation is the threshold selection, which usually results in a user bias. To compare the histograms of the HR-GT, EDSR-HM, and CinCGAN-HM images with the LR image, we cropped a subvolume (380x380x512 voxels) from the HR-GT, EDSR-HM and CinCGAN-HM images that had the same number of voxels as the LR image. Fig.5(a) shows the intensity histograms of the LR, HR-GT, EDSR-HM, and CinCGAN-images. It is clear that the LR, HR-GT, EDSR-HM, and CinCGAN-HM images have similar histograms, which means they can share similar thresholds for segmentation. This provides a comparative way to quantitatively appraise the physical accuracy of the images.

The optimal segmentation thresholds, however, cannot be resolved directly from the histograms provided in Fig.6(a) due to the wide intensity range between the main two peaks with relative high frequency. Herein, we calculated the image gradient magnitude map vs. voxel intensity for the HR-GT image, as shown in Fig.6(b), to determine the optimal thresholds. Regions of low gradient magnitude with high frequency indicate pure phases, which are macropores or grain while regions with high gradient magnitude are interfacial regions. We calculated intensity gradient magnitudes for the HR-GT image from twenty random interfacial regions. Then, the minimum gradient magnitude – 10.75 (intensity variation/voxel) was selected as the threshold of pure macro pore phases. Those regions under gradient magnitude of 10.75 are considered as pure macropore phase. In Fig.6(c), we extract a histogram for only those regions with gradient magnitudes between 0 and 10.75. The extracted histogram displays a clear separation between the macropores and grains. Therefore, we selected the optimal thresholds for watershed segmentation as 0-55 for macropore and 65-255 for grain, based on Fig.6(c). In addition, we also generated extra segmentations for the validation data by increasing/decreasing the optimal thresholds for sensitivity analyses.

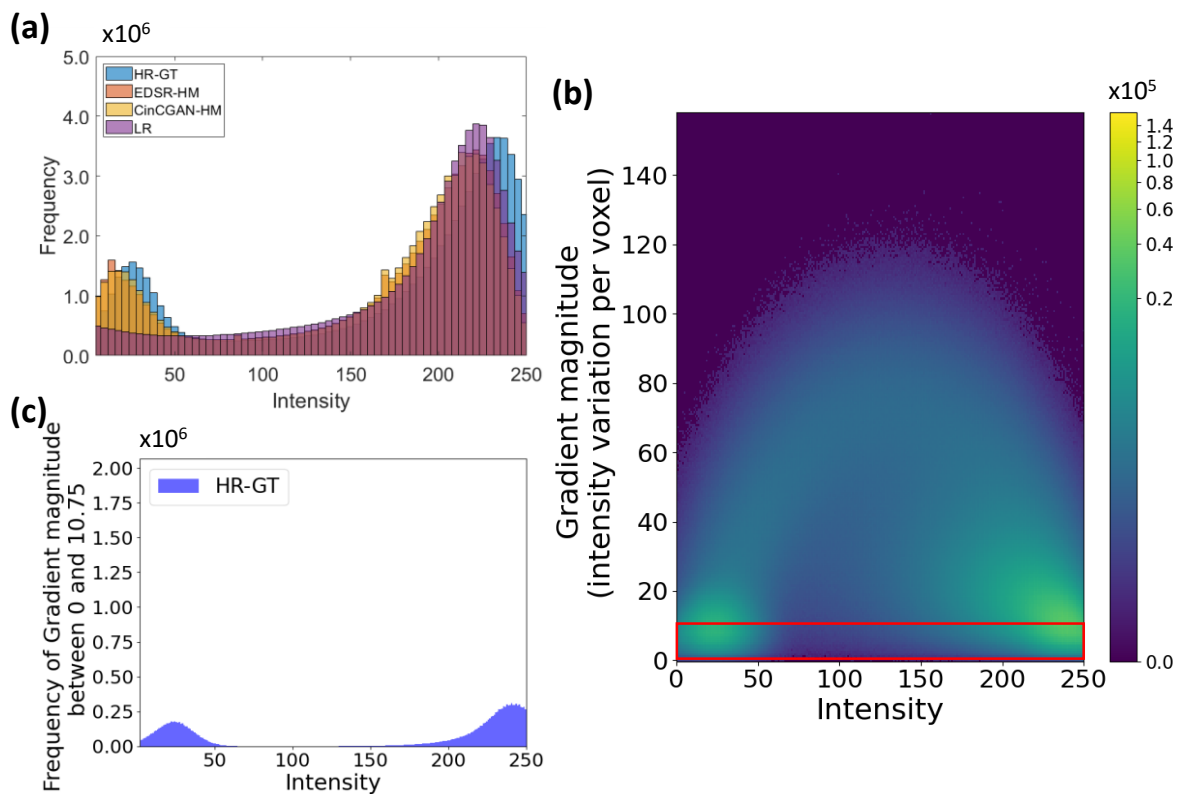

Fig.6: Optimal threshold determination on validation images for watershed segmentation. (a) Image intensity histograms of LR and HR-GT, EDSR-HM, CinCGAN-HM sub volumes (380x380x512 voxels), (b) Image intensity vs gradient magnitude histogram of HR-GT, (c) Image intensity vs gradient magnitude histogram of HR-GT between 0 and 10.75 intensity variation per voxel.

The differences between the EDSR-HM and CinCGAN-HM images can be observed in the segmented data (optimal thresholds). Fig.7(a)-(d) depict the 2D segmentations over many pores. Both the EDSR-HM and CinCGAN-HM images capture the finer macropores that are also captured in the HR-GT image. Region of interest (ROI) images are provided in Fig.7(e)-(h), which demonstrate that the EDSR-HM image can recover more representative pore structures than the CinCGAN-HM image, as noted by the blue box. Also, note that the segmented data presented in Fig.7 are taken from the same region as the grey-scale images presented in Fig.6.

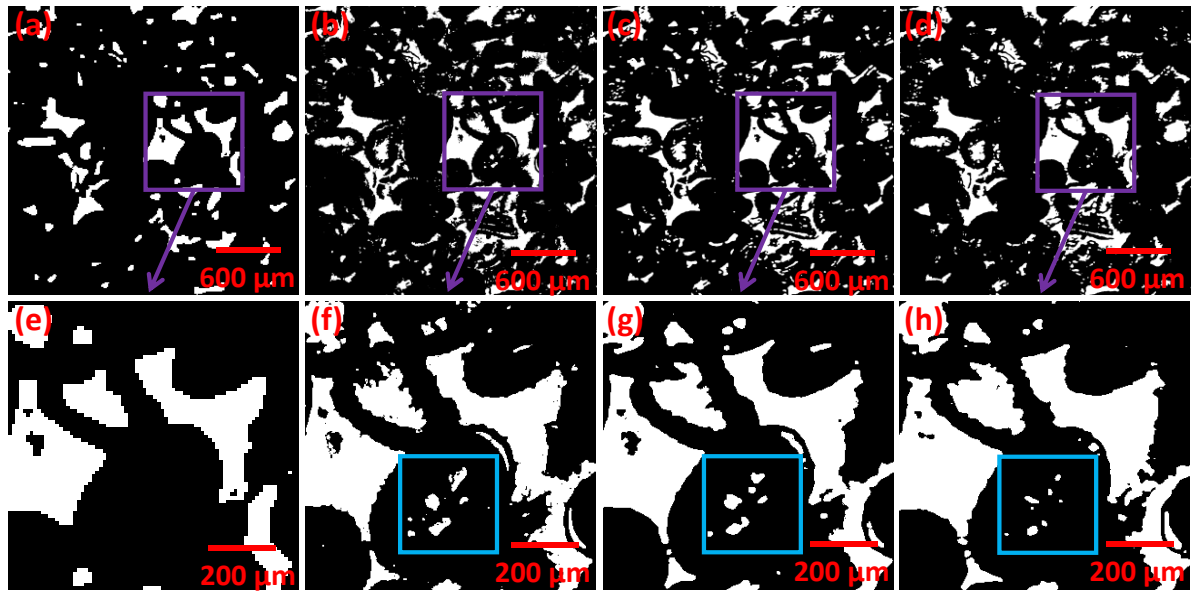

Fig.7: 2D optimal watershed segmentation with same FOV corresponding to Fig.7 (White: macro-pores, Black: grains). (a) LR, (b) HR-GT, (c) EDSR-HM, (d) CinCGAN-HM, (e) amplified LR, (f) amplified HR-GT, (g) amplified EDSR-HM, (h) amplified CinCGAN-HM.

### 3.3 3D Local Porosity Maps

To quantify the microporosity, we firstly multiply the grey scale image by the corresponding segmented image (Micro-pore:0, Grain:1) to obtain grey scale images with 'grain' phase only. A local porosity map for the grain phase region is generated by

$$\phi_{micro} = \frac{T_{(x,y,z)} - T_{grain}}{T_{pore} - T_{grain}}, \qquad Eq. 5$$

where $T_{(x,y,z)}$ is the intensity of the local position $(x, y, z)$ in the 3D image, $T_{pore}$ is the threshold of pure pore phase, $T_{grain}$ is the threshold of pure grain phase, and $\phi_{micro}$ is the range of micropore porosity between 0 and 1.

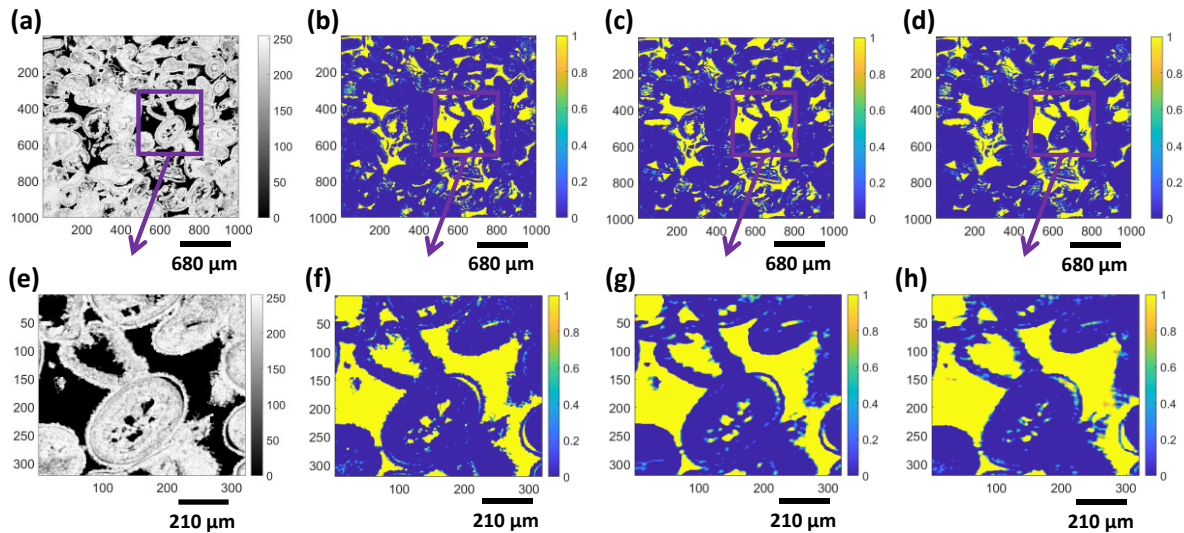

Fig.8: 2D local porosity map with same FOV corresponding to Fig.6 and Fig.7. (a) HR-HM grey scale image with grain phase, (b) local porosity map of HR-GT, (c) local porosity map of EDSR-HM, (d) local porosity map of CinCGAN-HM, (e) amplified HR-HM grey scale grain phase images, (f) amplified HR-GT grey scale image with grain phase, (g) amplified local porosity map of EDSR-HM, (h) amplified local porosity map of CinCGAN-HM. The color bars in sub-figure (a) and (e) represent the intensity range of the 8 bits grey-scale image. The color bars in the other sub-figures represent the range of the estimated micro-porosity related to Eq.5.

The generated the local porosity maps for the HR-GT, EDSR-HM, and CinCGAN-HM images are depicted in Fig.8. Firstly, in Fig.8(a)-(d), we observe that both EDSR-HM and CinCGAN-HM images can accurately recognise the microporosity as recognized in the HR-GT image. However, when observing the ROIs in Fig. 8(e)-(h), it is apparent that the EDSR-HM image in Fig.8(g) restores most of the microporosity characteristics compared with the HR-GT image while the CinCGAN-HM image shows a slightly larger fraction of microporosity.

Table 3 provides the macro/micro porosity values for the 3D validation volumes using the optimal segmentation thresholds. The porosity results show that both micro and macro porosities in the HR-GT image are representative of the MICP experimental data. The slight differences between MICP data and segmented data are caused by the segmentation error as well as the presence of heterogeneities since the validation images are not the same sample as used for MICP. In contrast, the LR image results in a large discrepancy when compared to the MICP data. The HR-GT image and MICP results show that our selected optimal thresholds are accurate enough to represent the geometrical information of the related volume.

Overall, our SR models provided relatively consistent porosity results. Compared with the HR-GT image and MICP data, the EDSR-HM image slightly overestimated macro/micro porosity while the CinCGAN-HM image slightly underestimated them. Overall, the results from the EDSR-HM and CinCGAN-HM images are close to the corresponding HR-GT images based on bulk micro and macro porosity. Additional porosity results on various segmentations can be found from S3-S4 in the Supplemental Material [65] in order to consider the uncertainty on the threshold values.

*Table 3: Macro/Micro porosity calculated on optimal segmentation of HR-GT, EDSR-HM, CinCGAN-HM, LR vs MICP experimental data [37].*

|  | MICP | HR-GT | EDSR-HM | CinCGAN-HM | LR |
|---|---|---|---|---|---|
| Macroporosity | 0.178 | 0.181 | 0.184 | 0.177 | 0.125 |
| Macroporosity error (%) | NA | 1.57 | 3.03 | -0.84 | -30.0 |
| Microporosity | 0.110 | 0.111 | 0.106 | 0.112 | 0.147 |
| Microporosity error (%) | NA | 1.37 | -3.65 | 2.46 | 33.91 |

In addition, to bulk porosity, we also investigate how the microporosity is distributed in the images. To quantify the microporosity distribution, we calculate the Dykstra-Parson coefficient curves proposed by Dykstra and Parsons, which is to measure the degree of heterogeneity [66] [67].

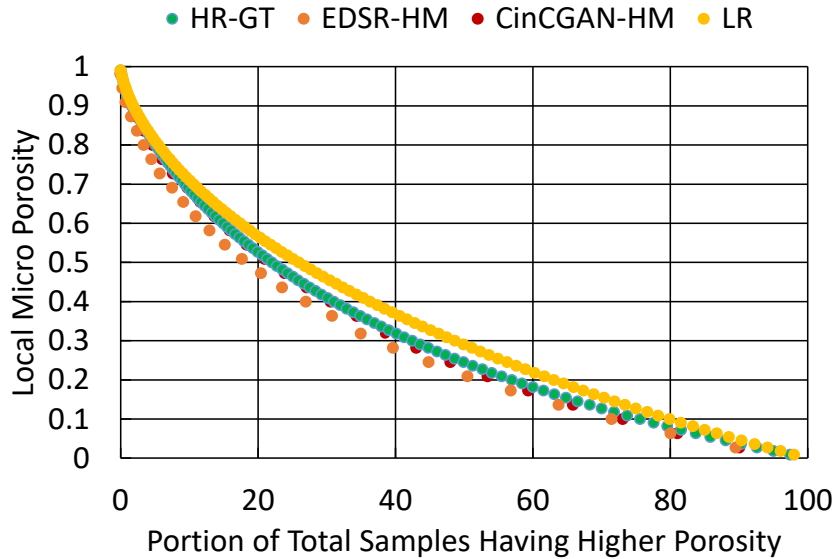

Fig.9: Dykstra-Parson coefficient for local micro porosity variation on HR-GT, EDSR-HM, CinCGAN-HM and LR.

Fig.9 shows the Dykstra-Parson coefficient curves for HR-GT, EDSR-HM, CinCGAN-HM and LR. Results show that the Dykstra-Parson coefficients of the CinCGAN-HM image are closer to the HR image than the EDSR-HM image. This indicates that CinCGAN can recreate the features of SR images that are comparable to HR-GT level. In addition, the Dykstra-Parson coefficient of LR generally have larger bias than HR-GT, EDSR-HM and CinCGAN-HM.

## 3.4 PNM for Petrophysical Analyses

The previous results showed that both the EDSR-HM and CinCGAN-HM images can resolve the macro/micro pores accurately compared with the HR-GT results. We further implement a PNM on the validation images to measure permeability, formation factor, capillary pressure, and relative permeability. The HR image is used as the ground truth data and the LR image is used as a baseline measure to assess the accuracy that is gained by using the SR algorithms.

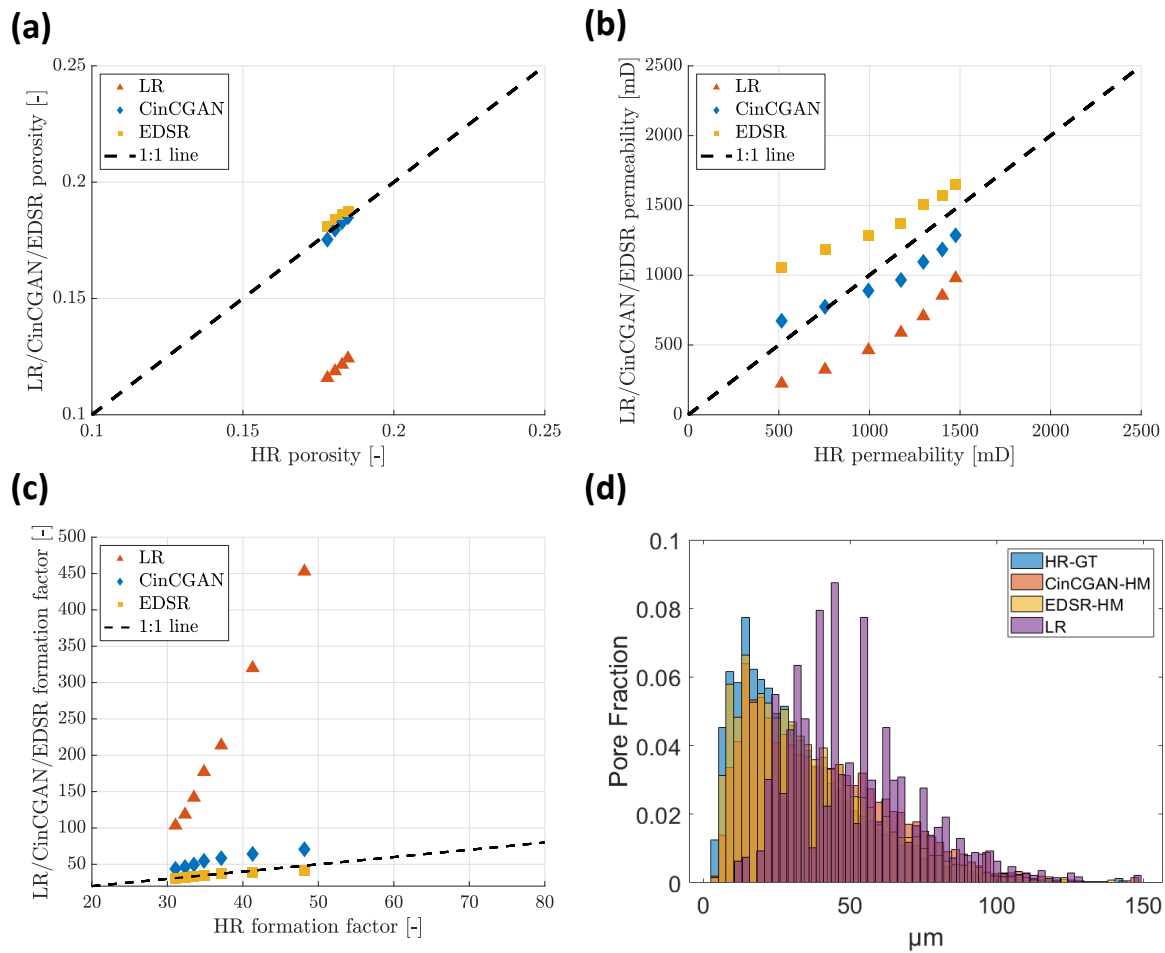

Fig.10 (a)-(c): Single phase/electrical flow PNM simulation of LR/HR-GT/EDSR-HM/CinCGAN-HM for absolute permeability, porosity and formation factor, d) pore size distribution for LR/HR-GT/EDSR-HM/CinCGAN-HM using local distance maximum method [68] [69] [70].

Fig.10 shows the single phase/electrical flow PNM simulations for the LR/HR-GT/EDSR-HM/CinCGAN-HM images. Each category contains seven segmentations with various thresholds around the pre-determined optimal. The HR results are represented on the X-axis as the benchmark. With thresholds increasing, porosity and absolute permeability increases appropriately in Fig.10 (a)-(b) for the LR/EDSR-HM/CinCGAN images. In general, the porosity results of both EDSR-HM and CinCGAN-HM are consistent with the HR-GT images within the tested threshold ranges. However, discrepancies can be found with the absolute permeability results where the CinCGAN-HM images had less deviation from the HR-GT images than the EDSR-HM images. Conversely, the formation factor results in Fig.10(c) show that the EDSR-HM results are more precise than the CinCGAN-HM results. In addition, all simulation results for the LR images do not correspond to the HR image results and display high variability over the tested thresholds. This is because the LR data has ambiguous boundaries between the pore and grain phases, as demonstrated by the high number of voxels that exist between the two main histogram peaks, see Fig.5(a).

Fig.10(d) shows the pore size distribution measured based on the local distance maximum method [68] [69] [70].The pore size distributions of the EDSR-HM and CinCGAN-HM images are mostly equivalent to the HR-GT image with only a few smaller pores resolved in the HR-GT images. Whereas the LR image resolves only larger pores and provides a limited range of pore size compared to the pore size distributions of the HR-GT, EDSR-HM, and CinCGAN-HM images. The LR image also provides many more large pores than the SR and HR counterparts suggesting that the pore space was over segmented when using the optimal thresholds.

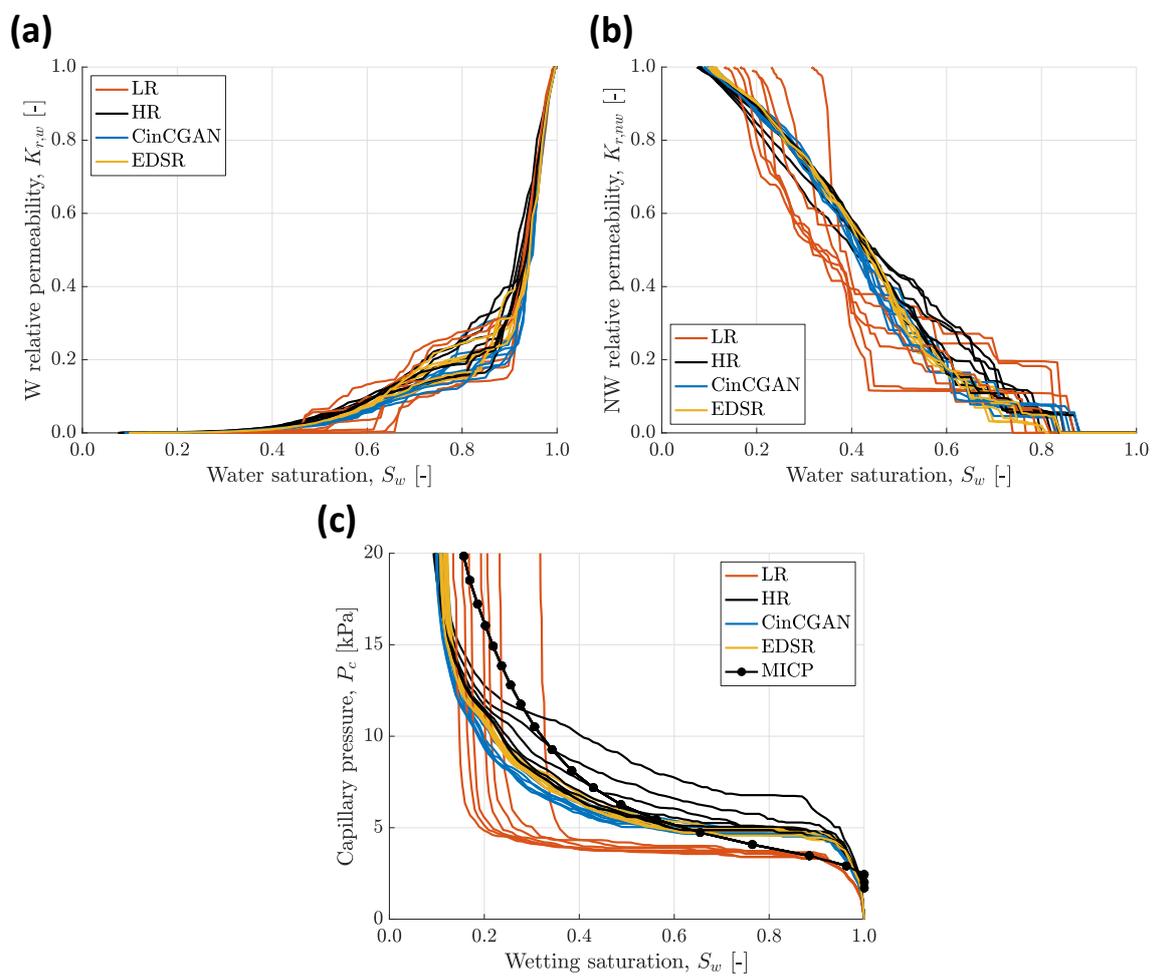

Fig.11: Multiphase flow PNM simulation of LR/HR-GT/EDSR-HM/CinCGAN-HM. a) non-wetting relative permeability, b) wetting relative permeability, c) drainage capillary pressure.

Fig. 11 shows the multiphase flow PNM results for the LR/HR-GT/EDSR-HM/CinCGAN-HM images. The relative permeability curves in Fig.11(a) demonstrate that both the EDSR-HM and CinCGAN-HM images are aligned smoothly with the HR-GT image while the LR image results are less correlated. Particularly in the specific range of $0.6 < S_w < 0.8$, the LR results show more non-continuous 'bounds' or 'step-like' features. This effect is more prominent in the non-wetting relative permeability curves shown in Fig. 11(b) while the EDSR-HM/CinCGAN-HM images are consistent with the HR results. As pore/throat sizes dominate how relative permeability varies across the water saturation range. These 'step-like' features indicate that there is a narrower variation of pore sizes in the LR images, which is consistent with the pore size distribution results shown in Fig.10d. In contrast, the relative permeability curves of the EDSR-HM, CinCGAN-HM, and HR-GT show smoother transitions across the entire saturation range. This indicates that EDSR-HM and CinCGAN-HM can resolve relatively small macropores comparable to the HR-GT PNM results while LR images only resolve the larger macropores.

The capillary pressure curves are shown in Fig.11(c). The capillary pressure variations in the EDSR-HM and CinCGAN-HM images are more consistent with the HR-GT images than the LR images. The capillary pressure curves for the LR images generally meet the irreducible water saturation point earlier than the EDSR-HM, CinCGAN-HM, and HR-GT counterparts. This indicates that less of the smaller macropores are resolved in the LR image. At a given saturation point, capillary pressure for the LR image is lower than the EDSR-HM and CinCGAN-HM images as well as the HR-GT images. This means that the resolved average pore sizes of the LR images are larger than the EDSR-HM, CinCGAN-HM, and HR-GT images. Overall, the HR-GT, EDSR-HM and CinCGAN-HM shows accurate correlations in the range of $0.5 < S_w < 1$ while the capillary pressure of HR-GT, EDSR-HM and CinCGAN are underestimated in the range of $S_w < 0.5$. This can be considered as a resolution restriction since MICP can detect more tiny micro pores than micro-CT data. More non-wetting phase fluids move to those micropores in the MICP experiment at low $S_w$. Consequently, the capillary pressure of non-wetting phase in MICP is larger than the capillary pressure estimated in HR images. This is why even HR results are still a bit off to the MICP data. It should also be noted that MICP was conducted on another core plug of the sample and not the same core plug as used for imaging.

In addition, when observing the relative permeability and capillary pressure in Fig.11, it becomes evident that both EDSR-HM and CinCGAN-HM are even less sensitive to the threshold variation than the HR-GT image. This effect actually reduced the user bias when determining the image segmentation settings. The effect occurs because the SR deep learning models utilise a quantization technique to reduce the model size and computational cost [71]. Consequently, the segmentations of the quantized grey-scale images in EDSR-HM and CinCGAN-HM have less noise and less intermediate grey-scale values, which subsequently reduced their sensitivity to threshold values.

## 4. Conclusion

A comparative study was proposed using paired and unpaired super resolution deep learning models for physically accurate digital rock images. A carbonate rock sample was scanned at low-resolution and 4x high resolution for EDSR and CinCGAN training. We then reconstructed an unseen low-resolution validation volume (380x380x512 voxels) to its super-resolution counterpart (1520x1520x2048 voxels) by EDSR and CinCGAN. A gradient-based method was implemented to select the optimal thresholds for image segmentation. Various segmentations were generated for macropores and grains around the optimal thresholds using a watershed-based method. The macroporosity and microporosity results obtained from the watershed segmentations are consistent with the HR image results as well as MICP experimental data.

*Table 4: Overall comparison of EDSR and CinCGAN performance using the optimal segmentation. The number of training data for both networks are similar, reconstruction time is estimated by the time cost of reconstructing a 1520x1520x2048 voxels SR volume from 380x380x512 voxels LR input validation data. The best performance value is in bold font.*

|  | EDSR | CinCGAN |
|---|---|---|
| Training Data Style | Paired | Unpaired |
| Total Training Time (mins) | 358 | **267** |
| Total Reconstruction Time (mins) | **8.5** | 17.2 |
| Macro-porosity error vs. MICP (%) | 3.03 | **-0.84** |
| Micro-porosity error vs. MICP (%) | -3.65 | **2.46** |
| Absolute Permeability error vs HR-GT (%) | **16.57** | -17.71 |
| Wetting phase relative permeability vs. HR-GT (%) | **-0.90** | -2.21 |
| Non-wetting phase relative permeability vs. HR-GT (%) | **-1.83** | -2.65 |
| Capillary Pressure vs. HR-GT (%) | -6.36 | **-6.26** |
| Capillary Pressure vs. MICP (%) | **-18.36** | -20.29 |

Furthermore, petrophysical properties were simulated using a PNM in a drainage water/decane system. Compared EDSR vs CinCGAN with the high-resolution ground truth data, petrophysical properties show that both the paired EDSR and unpaired CinCGAN methods can precisely restore the sharpness of the pores structures that are not well resolved in LR image. In addition, the petrophysical properties of the EDSR and CinCGAN images are equivalent to HR images through various segmentations while the LR image cannot represent the characteristics of the HR image.

Unlike EDSR which is a CNN-based approach by learning immediate mapping between LR and HR data, CinCGAN as a GAN-based method that aims to recreate realistic spatial features close to the distribution of real data. In other words, CinCGAN causes more uncertainty than EDSR since the realistic information is generated. Table 4 provides an overall performance comparison of EDSR and CinCGAN (The detailed methods of quantitative analyses can be found in Supplemental material [72]). Our results show that CinCGAN can generate realistic SR images that have equivalent performance to EDSR but requires 22.5% less computational time than EDSR when considering both training and reconstruction times. This means that unpaired GAN-based method is more flexible and less time-consuming than paired CNN method for real applications in digital rock.

Overall, we introduce an integrated workflow to enhance digital rock image resolution by examining the physical accuracy of paired and unpaired deep learning methods. Our results show that both paired EDSR and unpaired CinCGAN can reconstruct physically accurate SR images that are equivalent to the HR ground truth image. This unlocks new applications for using unpaired deep learning for digital rock image quality enhancement. The unpaired deep learning approach accelerates the application of SR methods since image registration is not required. Furthermore, decoupled digital rock data from retrieval platforms, such as the Digital Rock Portal (https://www.digitalrocksportal.org/), can be exploited more efficiently to deal with a wide range of geological data for image upscaling in a physically accurate way. Further studies of unpaired methods can be conducted for image resolution improvement of multimineral rock images, or other types of images collected from other imaging modalities, such as transmission electron microscopy, scanning electron microscopy, and X-ray computed tomography.

**Acknowledgement:**

The original micro-CT carbonate data in this paper is supported by digital rock portal.